
\input harvmac.tex
\Title{CTP/TAMU-10/93}{{Veneziano Amplitude for Winding Strings}
\footnote{$^\dagger$}{Work supported in part by NSF grant PHY-9106593.}}

\centerline{
Ramzi~R.~ Khuri\footnote{$^*$}{Supported by a World Laboratory Fellowship.}}
\bigskip\centerline{Center for Theoretical Physics}
\centerline{Texas A\&M University}\centerline{College Station, TX 77843}

\vskip .3in
String configurations with nonzero winding number describe soliton string
states. We compute the Veneziano amplitude for the scattering of arbitrary
winding states and show that in the large radius limit the strings always
scatter trivially and with no change in the individual winding numbers of the
strings. In this limit, then, these states scatter as true solitons.

\Date{3/93}

\lref\quartet{D.~J.~Gross,
J.~A.~Harvey, E.~J.~Martinec and R.~Rohm, Nucl. Phys. {\bf B256} (1985) 253.}

\lref\ccrk{C.~G.~Callan and R.~R.~Khuri, Phys. Lett. {\bf B261} (1991) 363.}

\lref\geo{R.~R.~Khuri, ``Geodesic Scattering of Solitonic Strings",
Texas A\&M preprint CTP/TAMU-79/92.}

\lref\scat{R.~R.~Khuri, ``Classical Dynamics of Macroscopic Strings",
Texas A\&M preprint CTP/TAMU-80/92 (to appear in Nucl. Phys. B).}

\lref\dghrr{A.~Dabholkar, G.~Gibbons, J.~A.~Harvey and F.~Ruiz Ruiz,
Nucl. Phys. {\bf B340} (1990) 33.}

\lref\dabhar{A.~Dabholkar and J.~A.~Harvey, Phys. Rev. Lett. {\bf 63} (1989)
478.}

\lref\mant{N.~S.~Manton, Phys. Lett. {\bf B110} (1982) 54.}

\lref\gsw{M.~B.~Green, J.~H.~Schwartz and E.~Witten,
{\it Superstring Theory} vol. 1, Cambridge University Press (1987).}

\lref\polch{J.~Polchinski, Phys. Lett. {\bf B209} (1988) 252.}

In recent work \scat, we obtained dynamical evidence for the
identification of the Dabholkar string soliton \refs{\dabhar,\dghrr}
with the underlying fundamental string by comparing the scattering of these
soliton solutions with expectations from a Veneziano amplitude computation for
macroscopic fundamental strings. These latter states were represented as $n=1$
winding states in the large winding radius limit. A computation of the
dynamical force between two identical strings \ccrk\ and of the metric on
moduli space for the scattering of two string solitons \geo\ both yielded the
result of trivial scattering, in agreement with the Veneziano amplitude
calculation.

In this paper we generalize the $n=1$ Veneziano amplitude result to arbitrary
incoming winding states. In particular, we find that the Veneziano
amplitude vanishes in the large radius limit except when the final
winding numbers are identical to the initial ones and the scattering
angle is zero. In other words, for arbitrary winding number, these states
scatter as true solitons.

The scattering problem is set up in four dimensions, as the kinematics
correspond essentially to a four dimensional scattering problem, and
strings in higher dimensions generically miss each other anyway \polch.
The precise compactification scheme is irrelevant to our purposes.

The winding state strings then reside in four spacetime dimensions
$(0123)$, with one of the dimensions, say $x_3$, taken to be periodic with
period $R$, called the winding radius. The winding number $n$ describes
the number of times the string wraps around the winding dimension:
\eqn\wind{x_3\equiv x_3 + 2\pi Rn,}
and the length of the string is given by $L=nR$. The integer $m$, called
the momentum number of the winding configuration, labels the allowed
momentum eigenvalues. The momentum in the winding direction is thus
given by
\eqn\pthree{p^3={m\over R}.}
The number $m$ is restricted to be an integer so that the quantum wave function
$e^{ip\cdot x}$ is single valued.
The total momentum of  each string can be written as the sum of a
right momentum and a left momentum \eqn\tmoment{p^\mu=p^\mu_R+p^\mu_L,}
where $p^\mu_{R,L}=(E,E\vec v,{m\over 2R}\pm nR)$,
$\vec v$ is the transverse velocity and $R$ is the winding radius.
The mode expansion of the general
configuration $X(\sigma,\tau)$ in the winding direction
satisfying the two-dimensional wave equation
and the closed string boundary conditions can be written as the sum of
right moving pieces and left moving pieces, each with the mode expansion
of an open string\gsw~
\eqn\movers{\eqalign{X(\sigma,\tau)&=X_R(\tau -\sigma) + X_L(\tau +\sigma)\cr
X_R(\tau -\sigma)&=x_R + p_R(\tau -\sigma) + {i\over 2}
\sum_{n=0} {1\over n}\alpha_n e^{-2in(\tau - \sigma)}\cr
X_L(\tau +\sigma)&=x_L + p_L(\tau +\sigma) + {i\over 2}
\sum_{n=0} {1\over n}\tilde\alpha_n e^{-2in(\tau + \sigma)}.\cr}}
The right moving and left moving components are then essentially
independent parts with corresponding vertex operators, number operators
and Virasoro conditions.

The winding configuration represented by $X(\sigma,\tau)$ describes a
soliton string state. It is therefore a natural choice to compare
the dynamics of these states with the Dabholkar string solitons
in order to determine whether we can identify the solutions of
the supergravity field equations with infinitely long fundamental strings.
Accordingly, we compared the scattering of $n=1$ winding states in
the limit of large winding radius with the scattering of the Dabholkar
solitons and in both situations found trivial scattering of identical
strings \scat. In this paper, we consider the dynamics of arbitrary winding
states.

Our kinematic setup is as follows. We consider the scattering of
two straight macroscopic strings in the CM frame with
winding numbers $n_1$ and $n_2$ and momentum number $\pm m$.
The incoming momenta in the CM frame are given by
\eqn\imoment{\eqalign{p^\mu_{1R,L}&=(E_1,E_1\vec v_1,{m\over 2R}\pm n_1R)\cr
p^\mu_{2R,L}&=(E_2,E_2\vec v_2,-{m\over 2R}\pm n_2R).\cr}}
The outgoing momenta (with momentum number $\pm m'$) are given by
\eqn\omoment{\eqalign{-p^\mu_{3R,L}&=(E_3,E_3\vec w_3,{m'\over 2R}\pm n_3R)\cr
-p^\mu_{4R,L}&=(E_4,E_4\vec w_4,-{m'\over 2R}\pm n_4R),\cr}}
where from conservation of momentum and winding number we have
\eqn\conso{\eqalign{&E_1+E_2=E_3+E_4\cr
&E_1\vec v_1+E_2\vec v_2=0 \cr
&E_3\vec w_3+E_4\vec w_4=0 \cr
&n_1+n_2=n_3+n_4\cr}}
and where $\vec v_i, i=1,2$ and $\vec w_k, k=3,4$ are the incoming and
outgoing velocities of the strings in the transverse $x-y$ plane.
For simplicity, assume $\vec v_1=v_1\hat x$ and
$\vec w_3=w_3(\cos\theta\hat x + \sin\theta\hat y)$.
For now we assume no longitudinal excitation ($m=m'$), but it turns out
that our analysis is unaffected by the possibility of excitation. In the
large $R$ limit, the transition amplitude for arbitrary longitudinal
excitation is dominated by a factor which decays exponentially with $R$.
Following the same counting argument as in \scat, one can show that the
number of possible excited transitions is bounded by a polynomial in
$R$. Thus the total amplitude is dominated by the exponential factor and it
is therefore sufficient to consider the $m=m'$ case in this limit.

As usual, the Virasoro conditions $L_0=\widetilde{L}_0=1$ must hold, where
\eqn\vops{\eqalign{L_0&=N+\half (p^\mu_R)^2\cr \widetilde{L}_0&=\widetilde
{N}+\half (p^\mu_L)^2 \cr}}
are the Virasoro operators\gsw\ and where $N$ and $\widetilde{N}$ are the
number operators for the right- and left-moving modes respectively:
\eqn\numbs{\eqalign{N&=\sum \alpha^\mu_{-n}\alpha_{n\mu}\cr
\widetilde{N}&=\sum \tilde\alpha^\mu_{-n}\tilde\alpha_{n\mu},\cr}}
where we sum over all dimensions, including the compactified ones.
It follows from the Virasoro conditions that
\eqn\evr{\eqalign{\widetilde{N}-N&=mn\cr
	E^2(1-v^2)&=2N-2+{({m\over 2R}+nR)}^2.\cr}}

We begin with a computation of the scattering of identical $n=1$ states but
with arbitrary final winding states (with total winding number adding up to
$2$). To that end, we set $n_1=n_2=1$ and consider for simplicity the
scattering
of tachyonic winding states. For our purposes, the nature of the string
winding states considered is irrelevant. A similar calculation for
massless bosonic strings or heterotic strings, for example, will be
slightly more complicated (involving kinematic factors), but will nevertheless
exhibit the same essential behaviour in the large radius limit (i.e.
exponential decay). For tachyonic winding states we have $N=\widetilde{N}=m=0$.
The kinematic setup reduces to
\eqn\kinset{\eqalign{p^\mu_{1R,L}&=(E,E\vec v,\pm R)\cr
p^\mu_{2R,L}&=(E,-E\vec v,\pm R)\cr
-p^\mu_{3R,L}&=(E',E'\vec w,\pm nR)\cr
-p^\mu_{4R,L}&=(2E-E',-E'\vec w,\pm (2-n)R),\cr}}
where conservation of momentum and winding number have been used.
\evr\ reduces to
\eqn\tevr{\eqalign{E^2(1-v^2)&=R^2-2 \cr E'^2(1-w^2)&=n^2R^2-2,\cr}}
with
\eqn\eprime{E'=E+{(n-1)R^2\over E}.}

In the standard computation of the four
point function for closed string tachyons, we rely on the independence
of the right and left moving open strings. For the tachyonic winding
state, we also separate the right and left movers with vertex operators
given by $V_R=e^{ip_R\cdot x_R}$ and $V_L=e^{ip_L\cdot x_L}$ respectively
to arrive at the following expression for the amplitude
\eqn\afour{A_4={\kappa^2\over 4}\int d\mu_4(z)\prod_{i<j}
|z_i-z_j|^{p_{iR}\cdot p_{jR}} |z_i-z_j|^{p_{iL}\cdot p_{jL}}.}
Since it easily follows from our kinematic setup that
$p_{iR}\cdot p_{jR}=p_{iL}\cdot p_{jL}$ holds
for this configuration, the tree level 4-point function
reduces to the usual Veneziano amplitude for closed tachyonic strings\polch
\eqn\veneziano{\eqalign{A_4&={\kappa^2\over 4} B(-1-s/2,-1-t/2,-1-u/2)\cr
&=({\kappa^2\over 4}) {\Gamma(-1-s/2)\Gamma(-1-t/2)\Gamma(-1-u/2)\over
\Gamma(2+s/2)\Gamma(2+t/2)\Gamma(2+u/2)},\cr}}
where the Mandelstam variables $(s,t,u)$ are identical for right and left
movers and are given by
\eqn\mandlestam{\eqalign{s&=4(E^2-R^2)\cr
t&=-2EE'(1+vw\cos\theta)+2nR^2-4\cr
u&=-2EE'(1-vw\cos\theta)+2nR^2-4.\cr}}
A quick check using \tevr\ and \eprime\ shows that $s+t+u=-8$.
For $n=1$, we recover the case considered in \scat. There we showed that
$A_4\to 0$ as $R\to\infty$ except for at the poles at $\theta=0, \pi$,
corresponding to trivial scattering for identical bosonic states.
We use the identity $\Gamma(1-a)\Gamma(a)\sin\pi a=\pi$ to rewrite $A_4$ as
\eqn\amptwo{A_4=({\kappa^2\over 4\pi})
\left[{\Gamma(-1-t/2)\Gamma(-1-u/2)\over \Gamma(2+s/2)}\right]^2
\left({\sin(-\pi t/2)\sin(-\pi u/2)\over\sin\pi s/2}\right).}
The sinusoidal factor contains the usual s-channel poles.
{}From the Stirling approximation $\Gamma(u)\sim\sqrt{2\pi}u^{u-1/2}e^{-u}$
for large $u$, we obtain in the limit $R\to\infty$
\eqn\ampthree{A_4\sim\left[{\alpha^\alpha \beta^\beta
\over \gamma^\gamma}\right]^2
\left({\sin(-\pi t/2)\sin(-\pi u/2)\over\sin\pi s/2}\right)}
where
\eqn\abc{\eqalign{\alpha&=EE'(1+vw\cos\theta)-nR^2\cr
\beta&=EE'(1-vw\cos\theta)-nR^2\cr
\gamma&=2(E^2-R^2).\cr}}
Note that $\alpha + \beta=\gamma$ and as a result the
exponential terms cancelled automatically. It also follows that $A_4$
reduces in the limit $R\to\infty$ to
\eqn\amp{A_4\sim\left({\alpha\over \gamma}\right)^{2\alpha}
\left({\beta\over \gamma}\right)^{2\beta}
\left({\sin(-\pi t/2)\sin(-\pi u/2)\over\sin\pi s/2}\right).}
It is easy to show that in the limit $R\to\infty$, $\alpha,\beta,\gamma\to
\infty$.
A tedious but straightforward computation using \tevr\ and \eprime\ shows that
$|\alpha/\gamma|\leq 1$ and $|\beta/\gamma|\leq 1$ with either equality (but
not both) being satisfied only for $n=1$ and $\theta=0,\pi$. In other
words for $n\neq 1$, $A_4\to 0$ exponentially as $R\to \infty$ for {\it all}
scattering angles. So the only possible final states are those with
$n_3=n_4=1$ and $\theta=0,\pi$. Note that for $n\neq 1$, there are no
poles in $A_4$ at $\theta=0,\pi$. Hence the 4-point function vanishes
exponentially with the winding radius away from the poles, which exist only
for $n=1$. As mentioned above, we can repeat the calculation
for $m'\neq m$, but we still get the same essential exponential decay
in the winding radius, with the number of possible excited transitions
bounded by a polynomial in $R$. In this limit, the amplitude is nonvanishing
only for $m'=m$, $n=1$ and $\theta=0,\pi$.

A similar situation can also be shown to arise in the case of oppositely
oriented strings. Going back to \imoment\ and \omoment, if we set
$n_1=-n_2=1$ and $n_3=-n_4=n$, then a similar calculation to the one above
shows that $A_4\to 0$ exponentially as $R\to\infty$
except when $n=1$ and $\theta=0$ (or $n=-1$ and $\theta=\pi$). Again, the
final states must be identical to the initial ones with zero scattering
angle. In particular, there is no annihilation in the large radius limit.
This would suggest that in a collision of oppositely oriented
Dabholkar string solitons, the strings would collide under the influence of the
attractive force between them but would emerge in the same final states.
One can go further and show that for $n_1=-n_2=n$ and $n_3=-n_4=n'$,
the amplitude vanishes in the large $R$ limit except when $n=n'$ and
$\theta=0$ or $n=-n'$ and $\theta=\pi$. The calculation in this case is
even more laborious, but is equally straightforward. Finally, one can
generalize to the case of arbitrary incoming winding states. Once more,
the amplitude is nonvanishing only when the final winding states are
identical to the initial ones and the scattering angle is zero. In fact,
the general case follows from the previous case by noting that since the
kinematics of the left and right movers decouple and we can essentially
consider one sector alone, the dynamics should not be affected by boosting,
say, the right movers to a frame in which we have opposite winding, whence
the trivial scattering result follows. In any event, the general case can
be explicitly shown, and, although it is considerably more tedious than the
simplest case, follows essentially the same line of argument.

The above calculations can be repeated for any other type of string, including
the heterotic string\quartet. The kinematics differ slightly from the tachyonic
case but the $4$-point functions are still dominated by an exponentially
vanishing factor in the large radius limit, and are nonvanishing only at
$\theta=0,\pi$ and when the final states are identical to the initial states.

The above anaylsis represented a tree-level computation in string theory.
It would be interesting to see whether the full quantum string loop
scattering amplitudes still yield trivial scattering for the macroscopic
winding states. In addition, it would be interesting to construct the
Dabholkar analogs for the higher winding states as well as their full
quantum string loop extensions.

\bigbreak\bigskip\bigskip\centerline{{\bf Acknowledgements}}\nobreak
I wish to thank Philip Rosenthal for helpful discussions.

\vfil\eject
\listrefs
\bye